\documentclass[12pt]{article}
\usepackage{graphicx,psfrag}
\usepackage{subfigure}
\usepackage{amsmath,amssymb}
\usepackage{type1cm}
\usepackage{bm}
\usepackage{comment}
\usepackage{mathptmx}
\usepackage{wrapfig}

\newcommand\beq{\begin{equation}}
\newcommand\eeq{\end{equation}}
\newcommand\beqa{\begin{eqnarray}}
\newcommand\eeqa{\end{eqnarray}}
\renewcommand{\d}{\partial}

\newcommand\bk{{\bf k}}
\newcommand\bq{{\bf q}}
\newcommand\E{\epsilon}

\newcommand\bp{{\bf p}}

\begin{document}
\title{Non-Fermi-liquid effect in magnetic susceptibility}

\author{T. Tatsumi and K. Sato}

\maketitle
\centerline{\it Department of Physics, Kyoto University, Kyoto 606-8502, Japan}

\begin{abstract}
Taking into account the
 anomalous self-energy for quarks due to the dynamic screening effect
 for the transverse gluon propagator, 
we study the temperature dependence of the magnetic susceptibility in detail. 
It is shown that there does not exist the $T\ln T$ term in the susceptibility, different
 from the specific heat, but an anomalous $T^2\ln T$ term arises instead  
as a novel non-Fermi-liquid effect. 
\end{abstract}

It is well known that Fermi liquid theory (FLT) is very powerful in
discussing the properties of interacting fermions at low temperature \cite{bay04,noz}.
The renormalization-group analysis have shown that the Fermi-liquid
theory is a fixed-point theory, where all the quasi-particle 
interactions are {\it marginal}
\cite{sha94}, except the attractive BCS channel. This argument, however,
 may be applied
to the case of the short-range interaction. 
Recent renormalization-group (RG) arguments have revealed 
that the quasi-particle interaction through the exchange of the
transverse gauge field is {\it relevant} and induces the non-Fermi-liquid
effects in gauge theories (QED/QCD) \cite{rei,gan,cha,nay,boy01,sch}.

Within FLT, fermions are treated as quasi-particles incorporating the 
self-energy; the quasi-particle interactions around the Fermi surface 
are important and physical quantities are given in terms of the
Landau-Migdal parameters.
In gauge theories (QED/QCD), there appear infrared (IR) singularities 
in the Landau-Migdal parameters due to their infinite range. 
To improve the IR behavior 
in the quasi-particle interaction, it is necessary to take into account the screening effect for
the gauge field. Actually, the screening effect have been shown to be 
important in the many-body theories \cite{noz}; the Coulomb
interaction becomes short-ranged by the Debye mass. 
The inclusion of the screening effect is also required
by the argument of the hard-dense-loop (HDL) resummation \cite{kap}. 
Anyhow we can 
see the static screening by the Debye
mass for the longitudinal mode and the IR behavior is surely improved. However, there is no
static screening for the transverse mode and there is only the dynamic 
screening \cite{kap}. 
Thus the IR singularities are still
left for the gauge interactions of the transverse mode.

Accordingly the self-energy of the quasi-particles,
$\Sigma_+(\E_\bk)$,     
exhibits an anomalous behavior as $\E_\bk\rightarrow \mu $ due to the
dynamic screening, 
${\rm Re}\Sigma_+(\E_\bk)\sim g^2/9\pi^2(\E_\bk-\mu ){\rm ln}(\Lambda/|\E_\bk- \mu |)$
within the one-loop calculation \cite{boy01,man} . 
Such an anomalous self-energy gives rise to the non-Fermi-liquid
behavior in entropy or specific heat \cite{rei,gan,cha,boy01,hol,ipp,ger05}.
An anomalous contribution to specific heat, $\propto T{\rm ln}T$ at low
$T$, has been firstly shown by Holstein et al. in the case of electron
gas\cite{hol}. 
It may be understood within FLT that the density of state at the Fermi
surface behaves like $\ln T$.
Recently the
analogous effect has been discovered in QCD \cite{boy01,ipp}. Similar effects due
to dynamical gauge fields in systems of strongly correlated electrons
were studied in refs. \cite{rei,gan,cha,nay}.  

In this Letter we study the magnetic susceptibility of the gauge
theories at finite temperature
\footnote{Here we only consider QCD, but our results are easily applied
to electron gas with small modification.}
. The magnetic susceptibility has been one of
the important physical quantities within FLT since the original work of
Landau, 
and repeatedly
utilized to study the magnetic properties in condensed-matter physics \cite{bay04,noz}.
Recently, the magnetic properties of QCD or its magnetic
instability would be an interesting subject \cite{tat00,nak03,nie05,ohn,
inu,pal,nam} 
in relation to phenomena of compact stars, especially
magnetars \cite{mag} or primordial
magnetic field in early universe, where one may expect the QCD phase
transition \cite{uni}.  

In a recent paper we have studied the magnetic susceptibility of quark
matter at $T=0$ within 
the FLT to figure out the screening effects for gluons on the
magnetic instability \cite{tat08, tat082}. 
We have seen that the transverse gluons 
still gives logarithmic singularities for the Landau-Migdal parameters,
but they cancel each other 
in the magnetic susceptibility to give a finite result. 

At $T\neq 0$, the Fermi surface is smeared over the width $O(T)$ , so 
that the dynamic screening effect should give rise to a logarithmic $T$ dependence 
for physical quantities.  Then, one may expect a similar non-Fermi liquid
effect in the magnetic susceptibility as in the specific heat, because
both quantities are related to the density
of states at the Fermi surface.
However, since there does not exist in the liquid 
the close relation between the specific heat and
the magnetic susceptibility that exists in gases, we shall see a
different non-Fermi-liquid effect.  
Actually we find 
that there appears $T^2\ln T$ term in the magnetic susceptibility as
another non-Fermi liquid effect.

In the following we consider the color-symmetric interaction among
quasi-particles: it can be written as the sum of two parts, 
the spin independent ($f^s_{\bk,\bq}$) and
dependent ($f^a_{\bk,\bq}$) ones;
\beq
f_{\bk\zeta,\bq\zeta'}=f^s_{\bk,\bq}+\zeta\zeta'f^a_{\bk,\bq}.
\label{qpint}
\eeq
Since quark matter is color singlet as a whole, the Fock exchange
interaction gives a leading contribution \cite{tat00,nak03,ohn}. We, hereafter,
consider the one-gluon-exchange interaction (OGE). 
Since the OGE interaction is a long-range force and we
consider the small energy-momentum transfer between quasi-particles, we
must treat the gluon propagator by taking into account the screening effects \cite{kap};
\beq
D_{\mu\nu}(k-q)=P^t_{\mu\nu}D_t(p)+P^l_{\mu\nu}D_l(p)-\xi\frac{p_\mu
p_\nu}{p^4} 
\eeq
with $p=k-q$, where $D_{t(l)}(p)=(p^2-\Pi_{t(l)})^{-1}$, and 
the last term represents the gauge dependence with a parameter
$\xi$. Note that the quasi-particle interaction
(\ref{qpint}) is independent of the gauge choice \cite{tat082}. 
$P^{t(l)}_{\mu\nu}$ is the standard projection operator onto the
transverse (longitudinal) mode \cite{kap}. 
Since the soft gluons should give a dominant contribution in our case,
we must sum up an infinite series of the polarization functions (the
hard-dense-loop (HDL) resummation) in the gluon propagator.
HDL resummation then gives the polarization functions for the transverse and longitudinal gluons 
as 
\beqa
\Pi_l(p_0,{\bf p})&=&\sum_{f=u,d,s} m_{D,f}^2,\nonumber\\
\Pi_t(p_0,{\bf p})&=&-i\sum_{f=u,d,s} \frac{\pi u_{F,f} m_{D,f}^2}{4}\frac{p_0}{|\bp|}, 
\eeqa
in the limit $p_0/|\bp|\rightarrow 0$, with $u_{F,f}\equiv
k_{F,f}/E_{F,f}$ and the Debye mass for each flavor,
$m_{D,f}^2\equiv g^2\mu_f k_{F,f}/2\pi^2$ \cite{kap}
\footnote{The Debye mass is given as $e^2\mu^2u_F/\pi^2$ for 
electron gas in QED.}
. 
Thus the longitudinal gluons are statically
screened to have the Debye mass, while the transverse gluons are
dynamically screened by the Landau damping. 
Accordingly, the screening effect for the transverse gluons is ineffective 
at $T=0$, where soft gluons contribute. At finite temperature, 
gluons with $p_0 \sim O(T)$ can contribute due to the diffuseness of the Fermi surface and the transverse gluons
are effectively screened.

We consider the magnetic
susceptibility at low temperature. 
We, hereafter, concentrate on one flavor and omit the flavor indices for
simplicity.
Magnetic susceptibility is then written in terms of the quasi-particle 
interaction~\cite{bay04,noz,tat082},
\beqa
\chi_M=\left(\frac{{\bar g}_D\mu_q}{2}\right)^2\frac{1}{\left.N^{-1}(T)+\bar
f^a\right.}
\label{chim}
\eeqa
where $\bar g_D$ is the gyromagnetic ratio \cite{tat082}.
$N(T)$ is an extension of the density of state at the Fermi surface for
$T\neq 0$; 
\beq
N(T)=-2N_c\int\frac{d^3k}{(2\pi)^3}\frac{\partial
n(\epsilon_\bk)}{\partial\epsilon_\bk}
\label{dos}
\eeq
with the Fermi-Dirac distribution function, 
$n(\epsilon_\bk)\equiv (1+e^{\beta(\epsilon_\bk-\mu )})^{-1}$,
where $\epsilon_\bk$ is the quasi-particle energy. At $T=0$ we have 
$N(0)=(N_ck_F^2/\pi^2)v_F^{-1}$ with the Fermi velocity,
$v_F=k_F/\mu-(N_ck_F^2/3\pi^2)f_1^s$ in terms of the Landau-Migdal
parameter $f_1^s$ \cite{bay76}.

$\bar f^a$ is a spin-dependent Landau-Migdal parameter given by 
\beq
{\bar f^a}\equiv -2N_c\int\frac{d^3k}{(2\pi)^3}\frac{\partial
n(\epsilon_\bk)}{\partial\epsilon_\bk}\int\frac{d\Omega_\bq}
{4\pi}\left.f^a_{\bk,\bq}\right|_{|\bq|=k_s}/N(T),
\label{fabar}
\eeq
where $k_s$ is defined by $\epsilon_{k_s}=\mu $ and coincides with the usual Fermi momentum $k_F$ at $T=0$. 

Note that in both Eqs.~(\ref{dos}) and (\ref{fabar}) the
function, $-\partial n(\epsilon_\bk)/\partial\epsilon_\bk$, is
sharply peaked at $\epsilon_\bk=\mu $ for $T/\mu \ll 1$, and we can see that only the
quasi-particles near the Fermi surface still gives a dominant
contribution. However, we cannot use the standard low-temperature
expansion, since the quasi-particle energy is not regular at the Fermi
surface due to the no screening for transverse gluons.

First we study the average of the density of state given by Eq.~(\ref{dos}).
We can rewrite it as
\beqa
N(T)&=&\frac{N_c}{\pi^2} \int_{\E_0}^{\infty} d\omega  \frac{d k}{d
\omega} k^2 \frac{\beta e^{\beta\left(\omega-\mu
\right)}}{\left(e^{\beta\left(\omega-\mu \right)}+1\right)^2}\nonumber\\
&\simeq& \frac{N_c}{\pi^2} \int_{\E_0}^{\infty} d\omega \left(1-
 \frac{\partial{\rm Re} \Sigma_+(\omega)}{\partial \omega}\right)k(\omega)E_{k(\omega)} \frac{\beta e^{\beta\left(\omega-\mu \right)}}{\left(e^{\beta\left(\omega-\mu \right)}+1\right)^2} \label{eq:NTomega}
  ,
\eeqa
with $\E_0 \equiv \E_{|\bk|=0}$, where $\omega$ is the quasi-particle
energy and $k(\omega)$ satisfies 
\beq
\omega = E_{k(\omega)} +{\rm Re}\Sigma_+(\omega,k(\omega)). \label{eq:omega}
\eeq 

The one-loop self-energy is almost independent of the momentum
\footnote{A
 renormalization group argument indicates that theory is infrared free in
 this case and the solution of the Schwinger-Dyson equation is almost
 the same as the one-loop result \cite{sch}.}
, and can be written
 as \cite{boy01,man}
\begin{align}
{\rm Re}\Sigma_+(\omega,k)\sim
{\rm Re} \Sigma_+(\mu ,k_F)&-\frac{C_fg^2u_{F}}{12\pi^2}(\omega-\mu )\ln\frac{\Lambda}{|\omega-\mu |} \notag\\
&+\Delta^{\rm reg}(\omega-\mu)
\label{eq:Sigma}
\end{align} 
around $\omega \sim \mu $ with $C_f=(N_c^2-1)/(2N_c)$ and $u_F=k_F/E_{k_F}$. $\Lambda$ is a
cut-off factor and should be taken as an order of the Debye mass, $\Lambda\sim
M_D\equiv \left(\sum_f m_{D,f}^2\right)^{1/2}$. 
The self-energy has an imaginary part, ${\rm Im}\Sigma_+(\omega,k)\sim
C_fg^2/24\pi|\omega-\mu|$, which measures the life time for
quasi-particles. In the following we only use the real part, since we
are interested in quasi-particle near the Fermi surface. 
Note that the
 anomalous term in Eq.~(\ref{eq:Sigma}) appears 
from the dynamic screening of the transverse gluons, while the
 contribution by the longitudinal gluons is summarized  
in the regular function $\Delta^{\rm reg}(\omega-\mu)$ of $O(g^2)$. The
longitudinal gluons then gives $O(g^2T^2)$ contribution to $N(T)$ as in
the usual situation.
Thus the leading-order contribution comes from the transverse gluons.
We, hereafter, extract only the transverse contribution, $N_t(T)$,
 using 
 the anomalous term in Eq.~(\ref{eq:Sigma}):
substituting Eq. (\ref{eq:Sigma}) into Eq. (\ref{eq:NTomega}), we
 obtain   
\begin{multline}
N_t(T)=\frac{N_ck_s\mu }{\pi^2}\Big[
 1+\frac{\pi^2}{6}\frac{(2k_F^2-m^2)}{k_F^4}T^2 \\
+\frac{C_fg^2u_F}{24}\frac{(2k_F^2-m^2)}{k_F^4}
T^2\ln\left(\frac{\Lambda}{T}\right)
 +\frac{C_fg^2u_F}{12\pi^2}\ln\left(\frac{\Lambda}{T}\right)
 \Big]+O(g^2T^2).
\label{eq:NT1}
\end{multline} 
$N_t(T)$ or its inverse, $N^{-1}_t(T)$, has a term proportional to $\ln
 T$ and gives the leading-order contribution. It has a singularity at
 $T=0$, which corresponds to 
the logarithmic divergence of the Landau-Migdal parameter $f_1^s$ at
 $T=0$ \cite{tat082,sat}.
We have kept the next-to-leading order term ($T^2\ln T$ term) in
 Eq.~(\ref{eq:NT1}), because we shall see that the $\ln T$ term is canceled out
 by another term appearing 
in the spin-dependent Landau-Migdal parameter $\bar f^a$ in the magnetic
 susceptibility
\footnote{This is a different feature from the specific heat, where $\ln
 T$ term remains in the final result as the leading-order contribution.} 
.
\begin{figure*}[h]
\begin{center}
\includegraphics[width=0.6\textwidth]{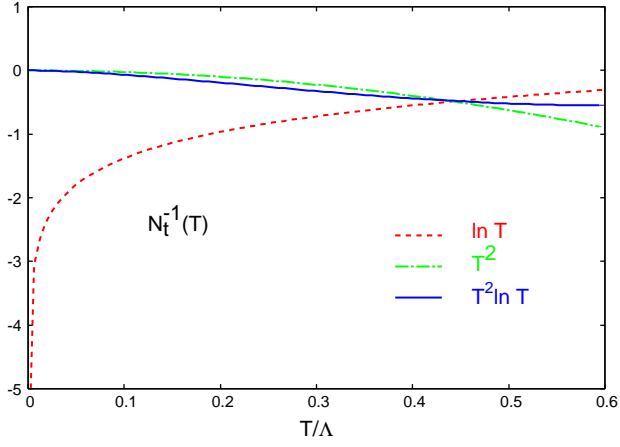}
\caption{Schematic view of each contribution to $N_t^{-1}(T)$. The
 leading order contribution ($\ln T$) is canceled in the magnetic
 susceptibility, so that the next-to-leading order contribution ($T^2\ln
 T$) becomes dominant.}
\end{center}
\end{figure*}

There are two contributions to $\bar f^a$: one is given by the
longitudinal mode, $\bar f^a_l$, and the other by the transverse mode, $\bar f^a_t$.
The transverse component $\bar f^a_t$ has a logarithmic singularity at
$T=0$ due
to the absence of the static screening. 
On the other hand, the longitudinal component $\bar f_{ l}^a$ has no IR singularity because
of the static screening, and is almost temperature independent. Thus 
the leading-order contribution at finite temperature comes from the
transverse gluons again as for $N(T)$. 
$\bar f_{ t}^a$ is given by
\begin{equation}
 \bar f_{ t}^a=2N_c N^{-1}(T)\int\frac{d^3k}{(2\pi)^3}\int\frac{d\Omega_\bq}{4\pi}
\frac{\partial n(\epsilon_\bk)}{\partial \epsilon_{\bk}} \left.\frac{m^2}{E_s
 E_\bk}C_fN_c^{-1}g^2M^{iia}(\bk,\bq) D_{ t}(k-q)\right|_{|\bq|=k_s}
\label{fat}
\end{equation}
where $M^{iia}$ is the reduced matrix element for the spin-dependent
interaction \cite{tat082}, and we defined $E_s$ by $E_s=E_{|\bq|=k_s}$.
It is the dynamic screening part in the propagator $D_t$ that gives the $\ln T$-dependence to $\bar f^a_{t}$. Therefore, we can put $|\bk|=k_s$ in the other parts of the integrand in Eq. (\ref{fat}).
The real part of the transverse propagator then render  
\begin{equation}
{\rm Re} D_{t}(k-q)\Big |_{|\bq|=k_s}\simeq -\frac{1}{2k_s^2}\frac{(1-\cos \theta_{\widehat{\bk\bq}})^2}{(1-\cos \theta_{\widehat{\bk\bq}})^3+ c^3(E_\bk-E_s)^2}
\end{equation} 
with  $c^3 \equiv
\frac{1}{8k_s^6}\sum_f\left(\frac{\pi m_{D,f}^2u_{F,f}}{4}\right)^2$,
while the imaginary part 
gives only higher order terms with respect to temperature and thus we neglect it here. 

For low $T$, the angular integrals in Eq.~(\ref{fat})  give 
a ${\rm ln}T$ dependence;
\beqa
\bar f_{t}^a \!\!\!\!\!\!&\simeq&\!\!\!\!\!\!
-\frac{C_fg^2}{12\pi^2 E_s^2T} N^{-1}(T)\!\!\!\int_{\E_0}^\infty \!\!\!d\omega
k(\omega)E_{k(\omega)}
\left(\!\!1\!-\!\frac{\partial {\rm Re}\Sigma_+(\omega)}{\partial \omega}\!\!\right)
\ln(\!|E_{k(\omega)}\!-\!E_s|\!)\frac{\partial n(\omega)}{\partial\omega}\nonumber\\
&\sim& -\frac{C_fg^2}{12N_c\mu^2}\ln T, \label{eq:fas}
\eeqa
where the term $\d {\rm Re}\Sigma_+(\omega)/\d\omega$ in the integrand does
 not contribute up to $O(g^2)$ in this calculation. Note that there appears
 no $T^2$ or $T^2\ln T$ term in the Landau-Migdal parameter.

Comparing Eq. (\ref{eq:fas}) with Eq. (\ref{eq:NT1}), one can see
that the $\ln T$ terms, which are the leading-order contribution,  exactly cancel each other in the magnetic
susceptibility (\ref{chim}) through the combination, $N^{-1}_t(T)+{\bar
f^a_t}$. Thus the remaining temperature
dependent terms in the
magnetic susceptibility have $T^2{\rm
ln}T$ terms as next-to-leading order (NLO) contribution, as well as
usual $T^2$ terms (see Fig.~1).

It is important to remember that the chemical potential is 
temperature dependent as well and other temperature dependence comes in the magnetic susceptibility.
The temperature dependence of the chemical potential can be derived by
considering the temperature variation on the quasi-particle number density
$\rho$, $d\rho/dT=0$ \cite{ipp}, 
\begin{equation}
\mu (T)=\mu _0-\frac{\pi^2}{6}\frac{(2k_F^2+m^2)}{ k_F^2E_F}T^2\left(1+\frac{C_fg^2}{12\pi^2}\ln \left(\frac{\Lambda}{T}\right)\right)\label{eq:muT}+O(g^2T^2).
\end{equation} 
It would be interesting to see that the chemical potential has $T^2\ln
T$ term besides the usual $T^2$ term due to the transverse gluons.
Taking into account this temperature-dependence in 
Eqs.~(\ref{eq:NT1}) and (\ref{eq:fas}), 
we finally find the temperature dependent part of the
magnetic susceptibility  $\delta\chi_M$,
\beqa
\delta\chi_M^{-1}=\chi_{\rm Pauli}^{-1}&\Big[&
\frac{\pi^2}{6k_F^4}\left(2E_F^2-m^2+\frac{m^4}{E_F^2}\right)T^2 
\nonumber\\
&+&\frac{C_fg^2u_F}{72k_F^4E_F^2} 
\left(2k_F^4+k_F^2m^2+m^4\right)T^2\ln\left(\frac{\Lambda}{T}\right)\Big]+O(g^2T^2),
\label{delT}
\eeqa
where $\chi_{\rm Pauli}$ is the Pauli paramagnetism, $\chi_{\rm
Pauli}\equiv {\bar g}_D^2\mu_q^2N_ck_F\mu/4\pi^2$.
It is evident that there appears $T^2 \ln T$ dependence in the
susceptibility 
at finite temperature besides the usual $T^2$ dependence.
This corresponds to $T \ln T$ term in the specific heat \cite{rei,gan,cha,boy01,hol,ipp,ger05} 
and is a novel {\it non-Fermi-liquid effect} in the magnetic susceptibility.
At low temperature, $\ln (\Lambda/T)>0$ so that the $T^2 \ln T$ term
gives positive contribution to 
$\chi_M^{-1}$.
Therefore, both $T$-dependent terms in Eq. (\ref{delT}) work against the
magnetic instability, which is characterized by the condition, $\chi_M\rightarrow 0$.

We have discussed the non-Fermi-liquid effect in 
the magnetic susceptibility for gauge theories (QED/QCD), 
where the screening effects for gluons are properly taken into account.
Since the quasi-particle energy is not regular on the Fermi surface, 
we cannot use the low temperature
expansion, different from the usual treatment in FLT.
Carefully extracting the temperature dependence, we have found that the
interesting
features of the magnetic properties in gauge theories, 
especially an anomalous $T^2 \ln T$ contribution 
to the magnetic susceptibility by the transverse gluons.
It may be interesting to recall that the static screening gives the
magnetic susceptibility  
the term proportional to $M_D^2
\ln M_D^{-1}$ at $T=0$ \cite{tat082}; the Debye mass $M_D$ works as an infrared (IR) cutoff to 
in the quasi-particle interaction due to the longitudinal gluons, while
there is no static screening for the transverse gluons. At
finite temperature, the Fermi surface is smeared over order $T$, 
so that temperature 
itself plays a role of the IR cutoff through the dynamic
screening in the
quasi-particle interaction due to the transverse gluons.   

The logarithmic temperature dependence appears in the magnetic susceptibility  
as a novel non-Fermi-liquid effect, and its origin is the same as in the well-known 
$T\ln T$ dependence of the specific heat~\cite{rei,gan,cha,boy01,hol,ipp,ger05}.   
However, recall that there is no relation between the specific heat and the
magnetic susceptibility within FLT, different from gases. Actually we have seen
that the ${\rm ln}T$ term in $N(T)$ is exactly canceled by the spin-dependent
interaction to leave $T^2{\rm ln}T$ term as a leading-order contribution.
The anomalous $T^2 \ln T$ term works against the magnetic
instability, as well as the usual $T^2$ term. 

We shall report elsewhere the consequences of the non-Fermi-liquid effect in more
detail and the magnetic phase diagram in QCD on the
density-temperature plane\cite{sat}.

This work was partially supported by the 
Grant-in-Aid for the Global COE Program 
``The Next Generation of Physics, Spun from Universality and Emergence''
from the Ministry of Education, Culture, Sports, Science and Technology
(MEXT) of Japan  and the Grant-in-Aid for Scientific Research (C) (16540246, 20540267).


\begin{thebibliography}{99}

\bibitem{bay04} G. Baym and C.J. Pethick, {\it Landau Fermi-Liquid
	Theory} (WILEY-VCH, 2004).\\
                A.A. Abrikosov, L.P. Gorkov and I.E. Dzaloshinski, {\it
	Methods of Quantum Field Theory in Statistical Physics}
	(Prentice-Hall. Inc., 1963).\\
                A.B. Migdal, {\it Theory of finite Fermi systems}
	(Intersci. Pub., 1967).

\bibitem{noz}P. Nozi\'eres, {\it Theory of Interacting Fermi Systems}
	 (Westview Press,1997).\\
            D. Pines and P. Nozi\'eres, {\it The Theory of Quantum
	Liquids} (Perseus books Pub., 1999).

\bibitem{sha94} R. Shanker, Rev. Mod. Phys. {\bf 66} (1994) 129.

\bibitem{rei} M.Yu. Reizer, Phys. Rev. {\bf B40} (1989) 11572; {\bf B44}
	(1991) 5476.

\bibitem{gan} J. Gan and E. Wong, Phys. Rev. Lett. {\bf 71} (1993) 4226.

\bibitem{cha} S. Chakravarty, R.E. Norton and O.F. Syljuasen, Phys. Rev. Lett. {\bf
	74} (1995) 1423.

\bibitem{nay} C. Nayak and F. Wilczek, Nucl. Phys. {\bf B430} (1994)
	534.\\
J. Polchinski, Nucl. Phys. {\bf B422} (1994) 617.

\bibitem{boy01} D. Boyanovsky and H.J. de Vega, Phys. Rev. {\bf D63}
	(2001) 034016; 114028.

\bibitem{sch} T. Sch\"afer and K. Schwenzer, Phys. Rev. {\bf D70} (2004) 054007; 114037.

\bibitem{kap} J.I. Kapusta, {\it Finite-temperature field
	theory} (Cambridge U. Press, 1993).\\
                M. Le Bellac, {\it Thermal Field Theory} (Cambridge
	U. Press, 1996).


\bibitem{man}C. Manuel and Le Bellac, Phys. Rev. {\bf D55}  (1997) 3215.\\
             C. Manuel, Phys. Rev. {\bf D62} (2000) 076009.


\bibitem{hol} T. Holstein, R.E. Norton and P. Pincus, Phys. Rev. {\bf
	B8} (1973) 2649.

\bibitem{bro} W.E. Brown, J.T. Liu and H. Ren, Phys. Rev. {\bf D61}
	(2000) 114012; {\bf D62}(2000) 054013.

\bibitem{wan} Q. Wang and D.H. Rischke, Phys. Rev. {\bf D65} (2002) 054005.

\bibitem{ipp} A. Ipp, A. Gerhold and A. Rebhan, Phys. Rev. {\bf D69}
	(2004) 011901.\\
              A. Gerhold, A. Ipp and A. Rebhan, Phys. Rev, {\bf D70} (2004) 105015; PoS (JHW2005) 013.

\bibitem{ger05} A. Gerhold and A. Rebhan, Phys. Rev. {\bf D71} (2005) 085010.

\bibitem{tat00} T. Tatsumi, Phys. Lett. {\bf B489} (2000) 280.\\
                T. Tatsumi, E. Nakano and K. Nawa, {\it Dark
	Matter}, p.39 (Nova Science Pub., New York, 2006).

\bibitem{nak03} E. Nakano, T. Maruyama and T. Tatsumi, Phys. Rev. {\bf
	D68} (2003) 105001.\\
                T. Tatsumi, E. Nakano and T. Maruyama,
	Prog. Theor. Phys. Suppl. {\bf 153} (2004) 190.\\
                T. Tatsumi, T. Maruyama and E. Nakano, {\it Superdense
	QCD Matter and Compact Stars}, p.241 (Springer, 2006).

\bibitem{nie05} A. Niegawa, Prog. Theor. Phys. {\bf 113} (2005) 581.


\bibitem{ohn} K. Ohnishi, M. Oka and S. Yasui, Phys. Rev. {\bf D76}
	(2007) 097501.

\bibitem{inu} M. Inui, H. Kohyama and A. Niegawa, arXiv:0709.2204.

\bibitem{pal} K. Pal, S. Biswas and A.K. Dutt-Mazumder, arXiv:0809.0404.

\bibitem{nam}
             S.-il Nam, H.-Y. Ryu, M.M. Musakhanov and H.-C. Kim, arXiv:0804.0056.

\bibitem{mag}
                P.M. Woods and C. Thompson, {\it Soft gamma ray
	repeaters and anomalous X-ray pulsars:magnetar candidates},
	Compact stellar X-ray sources, 2006, 547.\\
                A.K. Harding and D. Lai, Rep. Prog. Phys. {\bf 69}
	(2006) 2631.

\bibitem{uni}
            D. Boyanovsky, H.J. de Vega, D.J. Schwarz,
	Ann. Rev. Nucl. Part. Sci. (2006) 441.

\bibitem{tat08} T. Tatsumi, {\it Exotic States of Nuclear Matter} (World
	Sci., 2008) 272.

\bibitem{tat082} T. Tatsumi and K. Sato, Phys. Lett. {\bf B663} (2008) 322.

\bibitem{bay76} G. Baym and S.A. Chin, Nucl. Phys. {\bf A262} (1976)
	527.

\bibitem{sat} K. Sato and T. Tatsumi, to be submitted.





\end{thebibliography}
\end{document}